\documentclass{article}
\usepackage{spconf,amsmath,graphicx}

\usepackage{multirow}
\usepackage{makecell}
\usepackage[flushleft]{threeparttable}
\PassOptionsToPackage{hyphens}{url}\usepackage{hyperref}
\usepackage{soul,color}
\usepackage{booktabs}

\title{Improving Noise Robustness of an End-to-End Neural Model for Automatic Speech Recognition}
\name{\begin{tabular}{c}
		Jagadeesh Balam,
		Jocelyn Huang, 
		Vitaly Lavrukhin,
		Slyne Deng,
		Somshubra Majumdar,
	    Boris Ginsburg
	\end{tabular}}

\address{NVIDIA, Santa Clara, USA }

\usepackage{fancyhdr}
\fancypagestyle{pageStyleOne}{%
    \renewcommand{\headrulewidth}{0pt}
    \fancyhf{}
    \fancyfoot[L]{\scriptsize © 20XX IEEE. Personal use of this material is permitted. Permission from IEEE must be obtained for all other uses, in any current or future media, including reprinting/republishing this material for advertising or promotional purposes, creating new collective works, for resale or redistribution to servers or lists, or reuse of any copyrighted component of this work in other works.}
}

\begin{document}

\maketitle
\thispagestyle{pageStyleOne}

\begin{abstract}
We present our experiments in training robust to noise an end-to-end automatic speech recognition (ASR) model using intensive data augmentation. We explore the efficacy of fine-tuning a pre-trained model to improve noise robustness, and we find it to be a very efficient way to train for various noisy conditions, especially when the conditions in which the model will be used, are unknown. Starting with a model trained on clean data helps establish baseline performance on clean speech. We carefully fine-tune this model to both maintain the performance on clean speech, and improve the model accuracy in noisy conditions. With this schema, we trained robust to noise English and Mandarin ASR models on large public corpora. All described models and training recipes are open sourced in NeMo, a toolkit for conversational AI.
\end{abstract}

\section{Introduction}

Automatic Speech Recognition (ASR) systems based on end-to-end (E2E) neural networks (NN) architectures have been  setting new state-of-the-art (SOTA) for many speech benchmarks.  At the same time  publicly available speech corpora have been growing in size to thousands of hours \cite{panayotov2015librispeech,du2018aishell, mls2020}. NN-based models have the capacity to scale and learn from these huge speech datasets. In this paper, we describe  our experiments in building a large scale E2E ASR model exclusively on public datasets. Our goal is to train a highly accurate ASR model that is robust to various types of noises typically encountered in real-life applications.

This paper makes the following contributions:
\begin{enumerate}
  \item Presents fine-tuning with intensive data augmentation as an efficient method to improve noise robustness of an NN-based ASR model
  \item Demonstrates that a single ASR model trained with large and diverse speech dataset and then fine-tuned with noise augmentation performs well under various noisy conditions, handles multiple sample rates and different codecs 
  \item Gives rigorous evaluation of noise robust models against various noise sources 
 \item Provides open source models and training recipe for building a noise robust ASR for English and Mandarin languages using publicly available corpora 
\end{enumerate}

\section{Related Work}
Noise robustness of an ASR system has been an active research area and many different approaches have been proposed over time for both NN-based and Hidden Markov Model (HMM) based systems \cite{li2014overview}. For NN based systems, the approaches like front-end enhancements or model adaptation, data augmentation has been most effective in improving accuracy of the model~\cite{peddinti2015jhu,ko2017aug}.

Some excellent resources have been made available by the open source community for augmenting clean speech data with noise. With availability of diverse samples of real and simulated room impulse responses \cite{ko2017aug, szoke2019but} and various real world noise data \cite{font2013freesound} large scale augmentation that leads to very good generalization of ASR models is now possible. 
Efforts have also increased to provide more realistic datasets \cite{richey2018voices, szoke2019but} for noisy conditions to improve bench-marking and measuring research progress in this area. 

This work is most similar to noise robust large scale ASR models presented in \cite{amodei2016deep,narayanan2018domain}. Narayanan et al \cite{narayanan2018domain} showed that a single model trained on large diverse dataset generalizes well to various noisy conditions and can adapt much better to a new domain with very little data for fine-tuning. The main differentiation points of our contributions from prior work are the following: (1) to improve noise robustness of an E2E ASR model, we propose to use fine-tuning with noise augmentation; (2) all audio datasets in our work are publicly available. 

\section{Training and Fine-tuning ASR model with Noise Augmentation}

As an open source toolkit for conversational AI, NeMo \cite{kuchaiev2019nemo} comes with pre-trained models that are suitable for various applications. Our goal is to build a large scale model that generalizes well and can be used to support different applications. When an ASR system is built for a specific application many improvements can be made by trying to match the conditions with respect to the domain and the acoustic conditions. To build a model that performs well under various commonly occurring acoustic conditions like clean near-field speech, telephone environment and far-field scenarios, we first train a model  on ``clean''  datasets and then fine-tune it to improve accuracy under the following conditions:

\begin{enumerate}
\item Far-field speech
\item Low-rate speech codecs
\item Background noise
\item Mixed sample rate
\end{enumerate}
The larger scale of data used for these models makes saving multiple versions of augmented data not only cumbersome but also impractical. Instead, we use online data augmentation with noise and low rate codecs.

\subsection{Noise Augmentation}
For each utterance in the training batch, augmentation is applied with  probability $P_{aug}$ -- a hyper-parameter that controls the ratio of non-augmented data to augmented data used in training. For each augmented utterance the following steps are carried out:
\begin{enumerate}
\item With a probability $P_{rir}$, the audio signal is convolved with a random Reverberation Impulse Response (RIR)
\item Foreground noise added at random signal-to-noise-ratio (SNR) between $0$ and $30$ dB
\item Background noise added at an SNR randomly chosen between $10$ and $40$ dB
\end{enumerate}

\subsection{Codec Augmentation}
For codec augmentation, we choose AMR-NB -- a low rate speech codec widely used in mobile telephony and low rate settings of Ogg Vorbis codec \cite{moffitt2001ogg} to simulate distortions introduced by these codecs at low rates. We used Sound eXchange (SoX) library to encode and decode an utterance on-the-fly with one of the rates of these codecs. For AMR-NB, we randomly choose one of the lower five rates between 4.75 kbps and 7.4 kbps to simulate distortions introduced on narrow-band speech in telephony systems. To simulate low-rate codec losses on wide-band audio using the Ogg Vorbis codec, we choose lower rate settings from -1 to 4 where a setting of -1 encodes 16 kHz audio at around 25 kbps and a setting of 4 encodes audio at around 48 kbps.

\section {Datasets}

Our English model was trained on the following datasets: LibriSpeech \cite{panayotov2015librispeech}, Mozilla Common Voice \cite{ardila2019common}, Wall Street Journal (WSJ) \cite{paul1992design}, Fisher \cite{cieri2004fisher}, 
Switchboard \cite{godfrey1993switchboard} and NSC1-subset (read speech) from National Speech Corpus \cite{koh2019building} adding up to 6268 hours of training data. Audio from Fisher and Switchboard datasets was up-sampled to 16 kHz. After combining all the datasets, training data consisted of 3,923,934 utterances. Our Mandarin model was trained on AISHELL-2 dataset with 1000 hours of audio \cite{du2018aishell}. 

For noise augmentation, we construct a noise dataset using audio samples from the \textit{Freesound} database \cite{font2013freesound} and 
MUSAN corpus \cite{snyder2015musan}. We split the noise samples in chunks of 20 seconds. We use real and simulated room impulse responses (RIRs) from OpenSLR \cite{ko2017aug} and BUT ReverbDB \cite{szoke2019but}. Our noise training dataset consisted of 10,6465 noise segments and 66,101 room impulse responses.

\subsection{Test sets}
We evaluate our model on the following test sets:
\begin{enumerate}
    \item Far-field: retransmitted LibriSpeech test-clean from BUT ReverbDB
    \item Noise: noise added to Hub5'00 evaluation set (LDC2002S09, LDC2002T43) at various SNRs
    \item Codec noise: Hub5'00 evaluation set encoded and decoded with AMR-NB and LibriSpeech test-clean encoded and decoded with Ogg Vorbis codec
\end{enumerate}
Retransmitted LibriSpeech test-clean data from BUT ReverbDB \cite{szoke2019but} consists of test-clean data recorded in five different rooms. In each room 31 different microphones were used for recording audio. For our evaluation we used data from 3 microphones in four rooms (see Table \ref{tab:But_mics}). No data from these microphones was included in the training data.

{\renewcommand{\arraystretch}{1.0}
\begin{table}[ht]
\caption{Microphone positions used in far-field test sets from BUT ReverbDB.}
\label{tab:But_mics}
\centering
\scalebox{1.0}{
\begin{tabular}{c c c c} 
 \toprule
 \textbf{Room} & \textbf{Size, m$\times$m$\times$m} &  \textbf{Type} & \textbf{Test set Microphones}  \\
 \midrule
 Q301 & 10.7$\times$6.9$\times$2.6 & Office & 1, 15, 25\\
 L207 &	4.6$\times$6.9$\times$3.1 & Office & 1, 15, 31\\
 L212 & 7.5$\times$4.6$\times$3.1 & Office & 1, 15, 31\\
 D105 &	17.2$\times$22.8$\times$6.9 & Office & 1, 15, 31\\
\bottomrule
\end{tabular}
}
\end{table}
}

For simulated noise test sets we used noise data from VOiCES corpus \cite{richey2018voices} which includes babble noise, music, audio from television and background noise without any distractors recorded in far-field conditions in various rooms for different microphone positions.


\subsection{Training Methodology}
In all our experiments, we use a  QuartzNet-15x5 \cite{kriman2019quartznet} model trained with Connectionist Temporal Classification (CTC) loss \cite{graves2006}. QuartzNet employs 1D time-channel separable convolutions, a 1D version of depthwise separable convolutions \cite{howard2017mobilenets}. Each 1D time-channel separable convolution block can be separated into
a 1D convolutional layer with kernel $K$ that operates on each channel separately across $K$ time frames and a point-wise convolutional layer that operates on each time frame independently across all channels.

All audio is converted to $16$ kHz for training. After passing through the augmentation pipeline audio is processed with a front-end that outputs $64$ log mel spectrogram features calculated from 20 ms windows with a 10 ms overlap and which is then augmented with SpecCutout \cite{devries2017specutout}. 

We also found that a model trained with mixed-sample rates had negligible loss in accuracy compared to a model trained with a single sample-rate. For example, when we trained a model with only $8$ kHz data by resampling all the training data to 8 kHz, we did not see any improvement in accuracy on 8 kHz datasets compared to a $16$ kHz model trained with 8 kHz data upsampled to $16$ kHz along with data originally sampled at $16$ kHz or higher.

All models are trained with the NovoGrad optimizer \cite{novograd2019}, with $\beta_1 = 0.95$ and $\beta_2 = 0.5$ and a cosine-annealing learning rate policy. Table~\ref{tab:hyperparam_models} lists various hyper-parameters used for training. The models presented in this paper were trained on NVIDIA DGX-2 with mixed precision training. 
{\renewcommand{\arraystretch}{1.0}
\begin{table}[!h]
\caption{Training and fine-tuning parameters for QuartzNet English base (QN-En), noise robust (QN-En-NR), Mandarin base (QN-Mn) and noise robust QN-Mn-NR) models.}
\label{tab:hyperparam_models}
\centering
\scalebox{0.8}{
\begin{tabular}{c | c c | c c} 
 \toprule
 &\textbf{QN-En} & \textbf{QN-En-NR} &  \textbf{QN-Mn} & \textbf{QN-Mn-NR}  \\
 \midrule
 Language & English & English & Mandarin & Mandarin \\
  \midrule
 Data size (hrs) & 6268 & 6268 & 1000 & 1000  \\
 Fine-tuned? & N & Y  & N & Y \\
 Base Model & - & QN-En & - & QN-Mn \\
 Batch size & 32 & 32 & 32 & 32\\
 \# GPUs & 256 & 256 & 64 & 64\\
 Epochs & 600 & 200 & 400 & 200\\
 Intitial lr & 0.01 & 0.001 & 0.01 & 0.001 \\
 Warm-up steps & 1000 & 0 & 1000 & 0\\
 Weight decay & 0.001 & 0.001 & 0.001 & 0.001\\
 Noise $P_{aug}$ & 0 & 0.2 & 0 & 0.1\\
 CodecAug $p_{aug}$ & 0 & 0.1 & 0 & 0.5\\
 \bottomrule
\end{tabular}
}
\end{table}
}


\section{Results}
In this section, we present the results showing improvements in accuracy for various noisy test sets after fine tuning the base model with noise augmentation.
We use greedy WERs without the usage of an external language model throughout this paper to elucidate improvements solely due to  data augmentation used in acoustic modeling.
\subsection{Augmentation probability}
We first study the impact of augmentation probability $p_{aug}$ that controls the ratio of non noise-augmented data to noise-augmented data on WER of far-field BUT ReverbDB test sets and our ``clean" test sets. The base model was fine-tuned  with different probability $p_{aug}$ of augmentation and with $p_{rir}=1$. Table ~\ref{tab:Aug_prob} shows that the WER for clean test sets becomes worse with increasing augmentation probability and for $p_{aug}=1$ the model exhibits  drastic increase in WER for the clean test sets because the training examples used for fine-tuning only consist of RIR augmented utterances. This result agrees with earlier results presented in \cite{narayanan2018domain}, where the authors found that a value of 0.2 for the probability of augmentation gave them their best results.

{\renewcommand{\arraystretch}{1.0}
\begin{table}[!h]
\caption{The accuracy of QuartzNet English noise robust model (QN-En-NR) depending on augmentation probability, LibriSpeech and BUT ReverbDB test sets, greedy WER(\%).}
\label{tab:Aug_prob}
\centering
\scalebox{0.9}{
\begin{tabular}{c c c c c} 
 \toprule
 $\mathbf{P_{aug}}$ & \textbf{test-clean} & \textbf{test-other} & \textbf{D105} & \textbf{L207} \\
 \midrule
 0 & 3.86 & 10.05 & 29.22 & 34.45 \\
 0.3 & 4.03	& 10.37 & 18.42 & 22.82 \\
 0.5 & 4.14 & 10.58 & 17.72 & 22.13 \\
 0.7 & 4.33 & 10.9 & 17.72 & 22.13 \\
 1 & 72.91 & 55.78 & 17.45 & 21.88 \\
 
 \bottomrule
\end{tabular}
}
\end{table}
}

\subsection{Far-field Speech}
In Table~\ref{tab:WER_comparison_clean}, we list the WERs for clean and far-field speech for our English language models. In the first four rows, we show that the noise robust version of the model (QN-En-NR) is almost as accurate as the base model (QN-En) with negligible loss in accuracy for the worst case. For the far-field test sets we see up to  $66\%$ relative improvement  with QN-En-NR.

{\renewcommand{\arraystretch}{1.0}
\begin{table}[!h]
\caption{QuartzNet English base  (QN-En) vs  noise robust  (QN-En-NR), LibriSpeech and Hub5 Switchboard and CallHome, greedy WER(\%)}
\label{tab:WER_comparison_clean}
\centering
\scalebox{1.0}{
\begin{tabular}{c c c} 
 \toprule
 \textbf{Test set} & \textbf{QN-En} & \textbf{QN-En-NR} \\
 \midrule
 LS, test-clean & 3.86 & 3.98 \\
 LS, test-other & 10.05 & 10.11 \\
 Hub5'00 SWB & 10.2 & 10.3 \\
 Hub5'00 CHM & 16.2 & 16.2 \\
 \midrule
 \textbf{Far-field Test-clean} & &  \\
 \midrule
 D105 & 29.22 & 17.57 \\
 L207 & 34.45 & 21.53 \\
 L212 & 14.04 & 9.92 \\
 Q301 & 18.88 & 12.29 \\

 \bottomrule
\end{tabular}
}
\end{table}
}

\subsection{Mixed Sample Rate}
We show effectiveness of mixed sample rate training in Table ~\ref{tab:Mixed_rate }, where the original AISHELL-2 is sampled at 16 kHz. We used narrowband codecs AMR-NB and G.711 to encode and decode audio while training with $P_{aug}=0.5$ for this experiment. We again see that fine-tuning improves model accuracy on 8 kHz data with negligible difference on 16 kHz data.

{\renewcommand{\arraystretch}{1.0}
\begin{table}[!h]
\caption{QuartzNet-Mandarin base  (QN-Mn) and noise robust (QN-Mn-NR), AISHELL-2  $16$ and $8$ KHz, greedy WER(\%)}
\label{tab:Mixed_rate }
\centering
\scalebox{1.0}{
\begin{tabular}{c c c c} 
 \toprule
 \textbf{Test set}& \textbf{Sample freq, KHz} &\textbf{QN-Mn} & \textbf{QN-Mn-NR} \\
 \midrule
 dev\_ios & 16 & 6.86 & 7.15  \\
 test\_ios & 16 & 6.96 & 7.16  \\
  \midrule
 dev\_ios  & 8  & 12.35 & 7.48 \\
 test\_ios  & 8  & 12.25 & 7.52 \\
 \bottomrule
\end{tabular}
}
\end{table}
}

\subsection{Low rate codec noise}
Table~\ref{tab:WER_comparison_codec} shows robustness of QN-En-NR to distortions introduced by low rate codecs. We show the comparison for the lowest rates of Ogg Vorbis codec for 16 kHz speech and AMR-NB for narrowband speech.
{\renewcommand{\arraystretch}{1.0}
\begin{table}[!h]
\caption{QuartzNet-English base (QN-En) and noise robust model (QN-En-NR), LibriSpeech and Hub5 Switchboard and CallHome, greedy WER(\%)}
\label{tab:WER_comparison_codec}
\centering
\scalebox{0.9}{
\begin{tabular}{c c c c} 
 \toprule
  \textbf{Test set} &
  \textbf{Codec}&
  \textbf{QN-En} & \textbf{QN-En-NR} \\
 \midrule
 LS,test-clean & Ogg 25 kbps & 4.96 & 4.56 \\
 LS, test-other & Ogg 25 kbps & 11.99 & 13.73 \\
 Hub5'00 SWB & AMR-NB 4.75 kbps& 14.0 & 13.1 \\
 Hub5'00 CHM & AMR-NB 4.75 kbps & 22.1 & 19.9 \\

 \bottomrule
\end{tabular}
}
\end{table}
}

\subsection{Robustness to Noise}
\begin{figure}[htb!]
  \centering
  \includegraphics[width=1.0\linewidth]{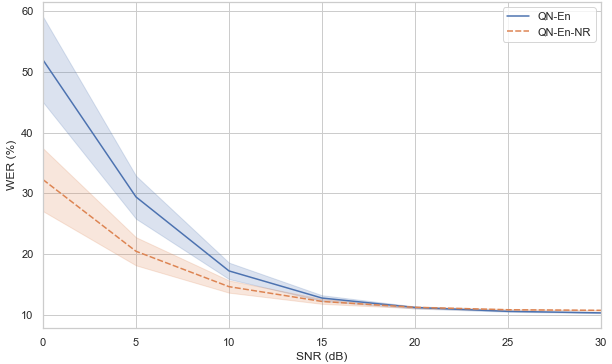}
  \caption{Quartznet-English base model (QN-En) and noise robust model( QN-En-NR) performance on noisy speech. Hub5 Switchboard eval set plus noise from VOiCES dataset, greedy WER(\%) averaged over 20 iterations.}
  \label{fig:hub5_noise_fig}
\end{figure}

\begin{figure}[htb!]
  \centering
  \includegraphics[width=1.0\linewidth]{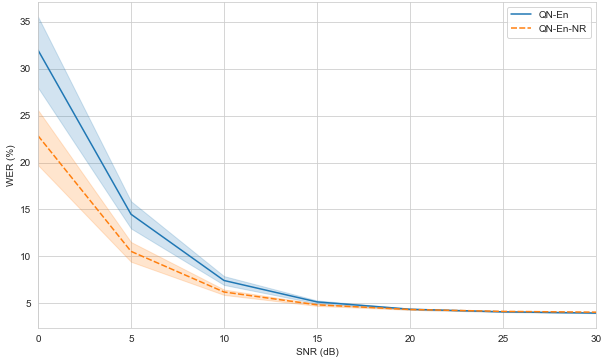}
  \caption{QuartzNet-English base (QN-En) and noise robust model (QN-En-NR) performance on noisy speech. LibriSpeech test-clean plus noise from VOiCES dataset, greedy WER(\%) averaged over 20 iterations.}

  \label{fig:libri_noise_fig}
\end{figure}

To evaluate our model's robustness to noise we simulate noisy datasets with noise added at various SNRs. For background noise we used retransmitted noise recordings from VOiCES corpus \cite{richey2018voices}. At each SNR, random noise segments are added to each utterance for 5 iterations for each of the four types of noises in VOiCES, resulting in 20 simulated noise sets at each SNR. In Fig.~\ref{fig:hub5_noise_fig} and Fig.~\ref{fig:libri_noise_fig}, we plot the
average WER for QN-En and QN-En-NR for test sets simulated with Hub5 SWB and LibriSpeech test-clean respectively. For narrow-band tests created with Hub5 SWB the noise segments derived from VOiCES were down-sampled to 8 kHz before being added to an utterance. Note the improved robustness to low SNR conditions with the noise robust model.

\section{Conclusions}
We showed that an E2E ASR model's noise robustness can be improved by fine-tuning the model with data augmentation. We first train the model on clean data to achieve best results on various clean data sets, then fine-tune this model with data augmentation. We presented results showing the efficacy of this method on English and Mandarin language models, and demonstrate that model trained with this method can both (1) generalize better to various acoustic conditions and also (2) maintain accuracy on the clean data set achieved with the base model.

The models are available at
\url{https://github.com/NVIDIA/NeMo}.



\bibliographystyle{IEEEbib}
\bibliography{strings, bibliography}

\end{document}